\documentclass[11pt,a4]{article}
\pdfoutput=1

\usepackage{amsmath}
\usepackage{epstopdf}
\usepackage{amsfonts}
\usepackage{amssymb}
\usepackage{graphicx, rotating}
\usepackage{epsfig}
\usepackage{latexsym}
\usepackage{mathrsfs}
\usepackage{graphicx}
\usepackage{color}
\usepackage{amsmath,amssymb}
\usepackage{cite}
\usepackage{slashed, cancel}
\usepackage{hyperref}
\usepackage{datetime}

\numberwithin{equation}{section}

\hypersetup{colorlinks, citecolor=bluscuro, linkcolor=bluscuro, urlcolor=bluscuro}
\definecolor{rossos}{rgb}{0.8,0.2,0.3}
\definecolor{bluscuro}{rgb}{0.15, 0.2, .85}
\definecolor{bluchiaro}{cmyk}{1,.3,0.,0.1}

\makeatother   

\def\d{{\rm d}}
\def\tr{{\rm tr\,}}

\newcommand{\MFNDi}{\mathcal{M}_1^{\rm FSR}}
\newcommand{\MINDi}{\mathcal{M}_1^{\rm ISR}}
\newcommand{\MLNDi}{\mathcal{M}_1^{(1)}}
\newcommand{\MTND}{\mathcal{M}^{(0)}}
\newcommand{\spinorubar}[1]{\bar{u}(#1)}
\newcommand{\spinorvbar}[1]{\bar{v}(#1)}
\newcommand{\spinoru}[1]{u(#1)}
\newcommand{\spinorv}[1]{v(#1)}

\def\be   {\begin{equation}}   \def\ee   {\end{equation}}
\def\ba   {\begin{array}}      \def\ea   {\end{array}}
\def\bea  {\begin{eqnarray}}   \def\eea  {\end{eqnarray}}
\def\bean {\begin{eqnarray*}}  \def\eean {\end{eqnarray*}}

\begin{document}

%
\begin{flushright} 
CERN-PH-TH/2013-108\\
SISSA  23/2013/FISI
\end{flushright}

\vspace{0.5cm}
\begin{center}

{\LARGE \bf The Role of Electroweak Corrections\\[0.4 cm] 
for the Dark Matter Relic Abundance}
\\[1.2cm] 
{\large\textsc{Paolo Ciafaloni} $^{\rm a}$,
 \textsc{Denis Comelli} $^{\rm b}$, 
 \textsc{Andrea De Simone} $^{\rm c,d,e}$,
 } \\[0.2cm]
{\large \textsc{Enrico Morgante} $^{\rm f}$,
  \textsc{Antonio Riotto} $^{\rm f}$, 
 \textsc{Alfredo Urbano} $^{\rm d}$}
\\[1cm]

\textit{$^{\rm a}$ Dipartimento di Fisica, Universit\`a di Lecce and INFN - Sezione di
Lecce, \\Via per Arnesano, I-73100 Lecce, Italy }

\textit{$^{\rm b}$  INFN - Sezione di Ferrara, Via Saragat 3, I-44100 Ferrara, Italy}

 \textit{$^{\rm c}$  CERN, PH-TH Division, CH-1211,
Gen\`eve 23,  Switzerland}

\textit{$^{\rm d}$ SISSA, Via Bonomea 265, I-34136 Trieste, Italy}

\textit{$^{\rm e}$ INFN, Sezione di Trieste, I-34136 Trieste, Italy}

\textit{$^{\rm f}$ D\'epartement de Physique Th\'eorique and Centre for Astroparticle Physics (CAP),}\\
\textit{24 quai E. Ansermet, CH-1211 Gen\`eve, Switzerland}
\end{center}

\vspace{0.5cm}

\begin{center}
\textbf{Abstract}
\begin{quote}
We  analyze the validity of the  theorems concerning  the  cancellation of the infrared and collinar divergences in the case of dark matter freeze-out in the early universe.
In particular,
we compute the electroweak logarithmic corrections of  infrared origin to the annihilation cross section of a dark matter particle being  the neutral component of a 
SU(2)$_L$ multiplet.
The inclusion of  processes with final state $W$ can modify significantly the
cross sections computed with only virtual $W$ exchange.
Our results show that the inclusion of infrared logs is necessary for a precise
computation of the dark matter relic abundance.
\end{quote}
\end{center}

\def\thefootnote{\arabic{footnote}}
\setcounter{footnote}{0}
\pagestyle{empty}

\newpage
\pagestyle{plain}
\setcounter{page}{1}

\section{Introduction}

The abundance of cold Dark Matter (DM) in our universe is now experimentally 
known with a great  precision, $\Omega_{\rm DM}h^2 = 0.1199\pm 0.0027$,  thanks to the Planck satellite recent data
\cite{Ade:2013zuv}. Very significant and diverse  experimental efforts are 
currently ongoing to identify
the nature and the interactions of the DM particle, other than 
the gravitational ones, but  in spite of this, a convincing signal is still missing.

A great deal of attention  has been recently devoted to the study of the effects
of ElectroWeak (EW) radiation, especially about their impact on the predictions for 
indirect DM searches \cite{Ciafaloni:2010qr, paper0, bell1, paper1, bell2, 
cheung, ibarra1, paper2, iengo, barger, ibarra2,proc, DeSimone:2013gj} (for earlier studies on the impact of
gauge boson radiation on DM annihilations or cosmic ray physics, see \cite{bergstrom1, bergstrom2, list1}). 

In this paper, we study the effects of EW corrections for DM
physics in a different setup, namely the epoch of DM freeze-out in the early universe,
when the DM relic density is established.
In view of the precision on the relic density measurements, several 
studies have been performed to refine the theoretical calculations at a level comparable
with the experimental uncertainties. For instance, the radiative one-loop corrections to the DM annihilation cross section have been computed, especially in the context of supersymmetric DM  (see e.g.~Refs.~\cite{Baro:2009na, Herrmann:2009mp,Boudjema:2011ig, Akcay:2012db,Harz:2012fz}). 
These studies are particularly
important in view of the fact that thermally  averaged cross section required to reproduce
the relic DM abundance can be calculated with a precision smaller than a few percent, 
once the relatively rapid changes (especially during the EW and QCD phase transitions) in the number of degrees of freedom with the temperature is properly accounted for
\cite{Steigman:2012nb}.

The radiation of  EW gauge bosons in the annihilation process are typically
not considered, relying on the cancellation of infrared and collinear divergences,
after a suitable average over the initial and final states is performed.
What we point out in this paper is that the conditions of validity of the cancellation
theorems are not satisfied if the DM particle belongs to an SU(2)$_L$ multiplet,
whose components are not degenerate in mass.
In such a situation, we find  potentially large corrections of infrared origin 
affecting the calculation of the annihilation cross sections, and hence the 
predictions for the DM relic abundance.

The paper is organized as follows. In Section \ref{sec:KLNviolation} we review
the cancellation theorems, their violations and we study the case of DM freeze-out.
In  Section \ref{sec:IRdiv} we present an explicit calculation of the infrared logs,
and show its relevance.
Finally, we summarize our main results and draw our conclusions in Section \ref{sec:conclusions}.
In  Appendix \ref{app:CollinearCancellation} we  explicitly work out 
the cancellation of  collinearly divergent logarithms, to show how the cancellation theorems work in the case of degenerate states.

\section{The violation of the Bloch-Nordsieck theorem for the DM freeze-out}
\label{sec:KLNviolation}

\subsection{The cancellation theorems and their violations}
 The Kinoshita-Lee-Nauenberg (KLN) theorem \cite{LN } states that in a field theory with massless fields 
the cross sections are free of both soft and collinear divergences if summed over
both final and initial degenerate states. 
Here degenerate states are meant to be  degenerate in energy  (infrared) and angle (collinear) up to
the resolving power of any given experiment. 
In fact, this cancellation theorem  for the InfraRed (IR) divergences in QED dates back to the pioneering work of Bloch and Nordsieck (BN) \cite{BN}  (see also \cite{coherent} for the coherent state approach) and successively extended to non abelian field theories \cite{ciafa}. 
The IR and collinear divergences manifest themselves as single or double logarithms of the ratio of the energy scale involved in the hard process over an IR scale and hence they can generate large radiative corrections.
 
We will show that potential large logs contributing to DM annihilation processes  are partially cancelled only once one defines appropriate averaged cross sections.
The precise relation between thermally-averaged cross sections and cancellation  theorems is our main result.

Let us assume that the DM particle is the electrically neutral Majorana component of a SU(2)$_L$ multiplet $\chi^a$ with hypercharge $Y=0$
 (for example, the case of a triplet can be the wino in supersymmetry). 
 Consider now the annihilation processes of the type $\chi^a\chi^b\rightarrow f$, $\chi^a\chi^b\rightarrow f+W_f$ (emission) and
 $\chi^a\chi^b+W_i\rightarrow f$ (absorption), where $f$ stands for a two-particle final state, typically made out of two fermions. 
Schematically, the  KLN theorem  states that the sum
\be
\label{klncompact}
\sum_{a,b}\sum_{i,f}  \left[
\left|{\cal M}^{(0)}+{\cal M}^{(1)}\right|^2_{ab\to f} +
 \int(\d\Pi_{W_f})\;\left|{\cal M}\right|^2_{ab\to f+W_f} + \int(\d\Pi_{W_i})\;\left|{\cal M}\right|^2_{ab+W_i\to f} \right]
\ee
is free of IR and collinear logarithmic divergences. Here 
 ${\cal M}^{(0)}$ and ${\cal M}^{(1)}$ are the tree-level and one-loop  amplitude  of the process
 $\chi^a\chi^b\rightarrow f$, respectively, and $\d\Pi_{W_{i,f}}$ indicate the integral measures for the 
 absorbed and emitted gauge boson, respectively.
In other words, the IR and collinear logarithmic divergences disappear upon averaging over the initial
flavor $a$ and the initial degenerate $W_i$'s in the initial states, and summing over the final flavor and  the final degenerate $W_f$'s in the final states. 
On the other hand,  the BN theorem \cite{ciafa}\footnote{The above cancellation mechanism  was shown to operate in the context of exact non-abelian gauge theories and later extended to the spontaneously broken abelian \cite{abelian} and non abelian field theories \cite{violation}, where different levels of inclusiveness are possible.} states that the sum
\be
\label{bncompact}
\sum_{a,b}\sum_{f}  \left[
\left|{\cal M}^{(0)}+{\cal M}^{(1)}\right|^2_{ab\to f} +
 \int(\d\Pi_{W_f})\;\left|{\cal M}\right|^2_{ab\to f+W_f}  \right]
\ee  
 is also free of IR divergences when one averages over the initial
flavors $a$  in the initial states and sums over the final flavor and the final degenerate $W_f$'s in the final states. 

If some of the hypotheses of these cancellation theorems does not hold,
their conclusions are evaded and in general 
 single or double logs (of infrared/collinear origin) will appear in the calculation 
 of the annihilation cross section. Therefore, when the cancellation theorems are not operative, 
one expects corrections  (at least) of the order 
\be
\label{estimate}
{\cal O}\left({\cal N}\cdot\frac{\alpha_W}{4\pi}\cdot\ln^n\frac{M_1^2}{m_W^2}\right)~,\qquad n=1,2\,,
\ee
where ${\cal N}$ is a numerical coefficient (typically large, as discussed later), $\alpha_W\equiv g^2/4\pi\approx 0.03$  and $M_1$  indicates the DM mass, assumed to be much larger than the weak scale
$M_1\gg m_W$.
Therefore, these corrections can be  large and even reach the order of 100\%. So, the crucial question is:  under which circumstances  are the cancellation theorems  not operative? 

One case in which this happens is at the LHC \cite{violation}. Indeed, since the colliding
initial states  carry a definite non-abelian flavor, EW radiative corrections to inclusive hard cross sections at the TeV scale are affected by peculiar BN-violating logs  \cite{violation, ew}.
 In fact, in such a case, the corrections are even the larger one in the estimate (\ref{estimate}) because the system is highly relativistic and  both collinear and IR divergences are present, leading to  double-log $\ln^2(M^2/m_W^2)$ enhancement.

Another case where large EW logs enter is in indirect searches for DM, 
where the relevant quantities are the energy spectra of final state particles originating from the annihilation of DM in the Milky Way.
These are partially inclusive quantities, as the energy spectra do not involve
integration over the full phase space of final particles, so the total inclusiveness
requested by the cancellation theorems is not respected.
Therefore,   the final fluxes of stable particles originated by DM annihilations are strongly affected by the real and virtual emission of $W$ gauge bosons. 

Finally, yet another situation where 
the cancellation theorems  fail to  screen the logarithmic corrections
is when the initial states are not degenerate in mass. In this case, the average over 
the initial flavor upsets the delicate cancellations of logarithms at work when the initial
state is completely mass-degenerate.
An example of this situation is the annihilation of DM particles, belonging to 
a SU(2)$_L$ multiplet, whose components are not degenerate in mass.
In the early universe,  around the epoch of the freeze-out of the DM particles when their relic abundance is established, 
the violation of the cancellation theorems is also of dynamical type. In order to make the theorems operative, one should sum over the initial states.
 However, they can be non-degenerate in mass and so differently populated in the thermal plasma and    differently weighted by an  exponential Boltzmann factor $\exp(-M_a/T)$, where $M_a\sim M_1$ is the mass of the
state $\chi^a$ and $T\sim M_1/25$ is the temperature of the universe around the freeze-out.
Conversely, only in the limit where all the particles of the multiplet which DM belongs to
are exacly degenerate in mass, the cancellation theorems apply and the correction (\ref{estimate}) is not present.
The study of this situation is  the central topic of the present paper.

\subsection{EW corrections for the DM freeze-out}\label{sec:EWfreeze}

Let us elaborate further about the violation of the cancellation theorems
in presence of non-degenerate DM multiplets and the impact this has on the
relic density calculation. 
 For definiteness, let us assume that the DM particle is the electrically neutral Majorana component of a SU(2)$_L$ triplet $\chi^a$ ($a=1,2,3$) with hypercharge $Y=0$.
 We will work out this example
 explicitly, but the conclusions are general and can be translated straighforwardly
 to any model where the DM is part of an SU(2)$_L$ multiplet.
  In order to capture the impact of the cancellation theorems in a model-independent way, we  work with an effective field theory Lagrangian, restricting ourselves to interactions of the DM triplet with the SM left-handed doublet $L=(f_1,f_2)^T$. The most general dimension-six operators are
 \be
 \mathscr{L}_{\rm eff}=\frac{C_{\rm D}}{\Lambda^2}\delta_{ab}\left(\overline{L}\gamma_\mu P_L L\right)
\left(\overline{\chi}^a\gamma^\mu\gamma_5  \chi^b\right) 
 +i\frac{C_{\rm  ND}}{\Lambda^2}\epsilon_{abc}\left(\overline{L}\gamma_\mu P_L \sigma^cL\right)
\left(\overline{\chi}^a\gamma^\mu  \chi^b\right)~,
\label{lagrangian}
 \ee
 where $C_{\rm D}$ and $C_{\rm  ND}$ are real coefficients for diagonal and non-diagonal 
 interactions in isospin space, $P_{R,L}=(1\pm\gamma_5)/2$ and $\sigma^c$ are the Pauli matrices. 

Let us consider the Boltzmann equation describing the
number density $n=\sum_a n_a$ of the particles $\chi^a$ in the plasma
\begin{eqnarray}
\label{boltz}
a^{-3}\frac{\d(n a^3)}{\d t}
&=&-\sum_{a,b,f}\langle v\sigma_{ab\rightarrow f} \rangle\left(n_a n_b- n_a^{\rm eq}n_b^{\rm eq}\right)\nonumber\\
&&-\sum_{a,b,f}\langle v\sigma_{ab\rightarrow f+W_f} \rangle\left(n_a n_b- n_a^{\rm eq}n_b^{\rm eq}\frac{n_{W_{f}}}{n^{\rm eq}_{W_{f}}}\right)\nonumber\\
&=&-\sum_{a,b,f}\left(\langle v\sigma_{ab\rightarrow f} \rangle+\langle v\sigma_{ab\rightarrow f+W_f} \rangle\right)\left(n_a n_b- n_a^{\rm eq}n_b^{\rm eq}\right)\,,
\end{eqnarray}
where in the last passage we have assumed that the gauge bosons are in kinetic equilibrium and we have disregarded the contribution from the absorption of the $W_i$, which in the Boltzmann equation is exponentially suppressed compared to the other terms. This is true
if we work in the regime $T\sim M_{1}/25 \lesssim m_W$, i.e. as long as the DM is lighter than a few TeV. 
Under this approximation of neglecting the absorption of $W$ from the initial state, 
the cancellation theorem possibly at work here is the BN theorem, and evading this
theorem would give rise to IR logs.

Upon introducing the quantities \cite{three}
\be
r_a=\frac{n_a^{\rm eq}}{n^{\rm eq}}=\frac{g_a(1+\Delta_a)^{3/2}e^{-x\Delta_a}}{g_{\rm eff}}~,
\ee
where $x=M_1/T$, $\Delta_a=(M_a-M_1)/M_1$ and $g_{\rm eff}=\sum_a g_a(1+\Delta_a)^{3/2}{\rm exp}(-x\Delta_a)$, one can rewrite the Eq. (\ref{boltz}) as
\be
a^{-3}\frac{\d(n a^3)}{\d t}=-\langle v\sigma_{\rm eff} \rangle\left(n^2- n_{\rm eq}^2\right)~,
\ee
where
\be
\sigma_{\rm eff} =\sum_{a,b,f} r_a r_b\left(\sigma_{ab\rightarrow f} +\sigma_{ab\rightarrow f+W_f}\right)~.
\ee
This object almost contains the expression (\ref{bncompact}), with the crucial difference of the weights $r_a$
\be
\sum_{a,b}\sum_{f} r_a r_b \left[
\left|{\cal M}^{(0)}+{\cal M}^{(1)}\right|^2_{ab\to f} +
 \int(\d\Pi_{W_f})\;\left|{\cal M}\right|^2_{ab\to f+W_f}  \right]\,.
\ee  
An exactly mass-degenerate DM multiplet would have the same $r_a$, for all $a$,
and the expression (\ref{bncompact}) would be recovered.
Instead, if the masses of the $\chi^a$ particles are different 
from each other, the BN theorem is not operative. In particular, if the mass splitting
\be
\Delta_a\gtrsim \frac{1}{25}~,
\ee
(where $x_f=M_1/T_f\simeq 25$ and $e^{-x_f\;\Delta_a}\leq 1$) we expect logarithmic divergences of the form (\ref{estimate}) to appear in the  computation of the 
final abundance. Notice that in the cosmological setup only single logs of IR origin appear, as the initial states are slightly  non-relativistic. The reader can find more details about this point in the Appendix \ref{app:CollinearCancellation}.

Conversely, in the regime of efficient annihilation $\Delta_a\lesssim 1/25$ (where $e^{-x_f\;\Delta_a}\sim 1-x_f\;\Delta_a$),  we expect that the common computation of the freeze-out abundance of DM is correct up to corrections at most of the order
\be\label{eq:ResidualKLNviolation}
{\cal O}\left(25\cdot\Delta_a\cdot{\cal N}\cdot\frac{\alpha_W}{4\pi}\cdot\ln\frac{M_1^2}{m_W^2}
\right)~.
\ee
The $2 \to 2$ cross sections include the EW one loop radiative corrections and are affected by Sudakov form factor corrections of single and double log size coming mainly from the final state virtual corrections \cite{sudakov}.
The leading double-log corrections   are then cancelled by the inclusion of real $W$ final states that reduce the degree of singularity to single-log factors whose origin can be identified with a violation of the BN theorem \cite{violation} and the fact that the initial states are partially  non-relativistic.
Finally, the fact that the DM multiplet contains particles almost mass-degenerate during the cohannilation processes  partially screens the single logs by a further coefficient $\Delta_a$ \cite{anomalous}.
In Table \ref{table1} we  give schematically the order of magnitude of the IR corrections expected once we compute
different kinds of cross sections involved  in the evaluation of the thermal relic abundance for different levels of inclusiveness. 
\begin{table}[t!]
\centering
\begin{tabular}{||c|c||}\hline
\textbf{Cross Section} & \textbf{Expected Corrections} \\ \hline\hline
$2\to 2$ with 1-loop radiative corrections & $\alpha_W \ln\frac{M_1^2}{m_W^2}, \alpha_W \ln^2\frac{M_1^2}{m_W^2}$ \\ \hline
$2\to 2 +W_f$, no average over initial flavor
 & $\alpha_W \ln\frac{M_1^2}{m_W^2} $ \\ \hline
$2\to 2 +W_f$,  average over initial non-degen. flavor  & $\alpha_W\;\Delta_a\; 
\ln\frac{M_1^2}{m_W^2}$ \\ \hline
\end{tabular}
\caption{{\small\textit{EW IR corrections involved into the evaluation of the various cross sections depending on the degree of inclusiveness. 
The first line refers to the usual Sudakov logs appearing when  virtual corrections
are included.
The correction in the second line appears when no average over the initial state
is performed, and BN theorem is evaded.
The last correction
originates from for the dynamical  violation of the IR cancellation provided by the BN theorem in thermal averaged cross sections, 
in presence of non-degenerate initial states.}}}
\label{table1}
\end{table}

We  stress also that the inclusion, beyond the $2\to 2$ processes, of the
 $2 \to 2+W_f$ ones, can induce a drastic modification of the cross sections involved into the annihilation processes.
In the case of the Lagrangian in Eq.~(\ref{lagrangian}), for example,  $2\to 2$ processes are proportional always to $|C_{\rm D}|^2$
while the  $2 \to 2+W_f$ are proportional to  $|C_{\rm ND}|^2$.
 Such a structural effects are generically included in the coefficient ${\cal N}\propto |C_{\rm D}/C_{\rm ND}|^2$ that  is potentially very large.

\section{An explicit example  of IR divergences}
\label{sec:IRdiv}

In this Section we analyze the relevance of the cancellation theorems from a more quantitative point of view. For this purpose, we shall retrace the arguments presented in Section~\ref{sec:KLNviolation} considering the effective Lagrangian in Eq.~(\ref{lagrangian}) as a benchmark model for the explicit calculations. Moreover - in order to catch in the simplest way the relevance of the IR corrections - we use throughout this section the eikonal approximation. This approximation, in fact, controls the long-range dynamic of the soft gauge boson emission thus providing the best-suited tool for the analysis of the IR singularities. In more details, the key point of the eikonal approximation is that the exchange of a soft gauge boson with four-momentum $k$ between two external legs with four-momenta $k_{1,2}$ is described by the following integral
\begin{equation}\label{eq:EikonalIntegral}
\mathcal{L}_{W}=-\frac{g^2}{2}\int
\frac{d^3k}{(2\pi)^32E_{k}}
\left(
\frac{k_1^{\mu}}{k\cdot k_1}-\frac{k_2^{\mu}}{k\cdot k_2}
\right)^2~.
\end{equation}
If the two external legs in Eq.~(\ref{eq:EikonalIntegral}) are both ultra-relativistic, then the value of the eikonal integral displays the well-known \cite{sudakov, violation} double logarithmic behavior $\mathcal{L}_{W}=
(\alpha_{W}/4\pi)\ln^2(s/m_W^2)
$; in this kinematical configuration, in fact, both the soft (i.e. when the energy of the emitted gauge boson approaches zero) and the collinear (i.e. when the momentum of the emitted gauge boson becomes parallel to the direction of the emitting particle) singularities are available in the phase space. This simple picture, however, is no longer true for the calculation of the relic density, where we have to deal with a different kinematic involving non-relativistic DM particles  ($v^2\sim 1/4$, being $v$ the DM relative velocity).  In this case the exchange of a gauge boson in the initial state produces the eikonal factor  
\begin{equation}\label{eq:EikonalNonRel}
\mathcal{L}_W=\frac{\alpha_W}{4\pi}\,\frac{4}{3}\,v^2\,\ln\frac{M_1^2}{m_W^2}~.
\end{equation} 
The resulting single logarithm in Eq.~(\ref{eq:EikonalNonRel}) is related to the residual soft singularity, while the collinear emission is kinematically forbidden. The interested reader can find more details about this point in Appendix~\ref{app:CollinearCancellation}. Notice that the result in Eq.~(\ref{eq:EikonalNonRel}) is proportional to the DM relative velocity $v$ because a particle at rest does not emit.

\begin{figure}[t]
\begin{center}
\includegraphics[scale=0.75]{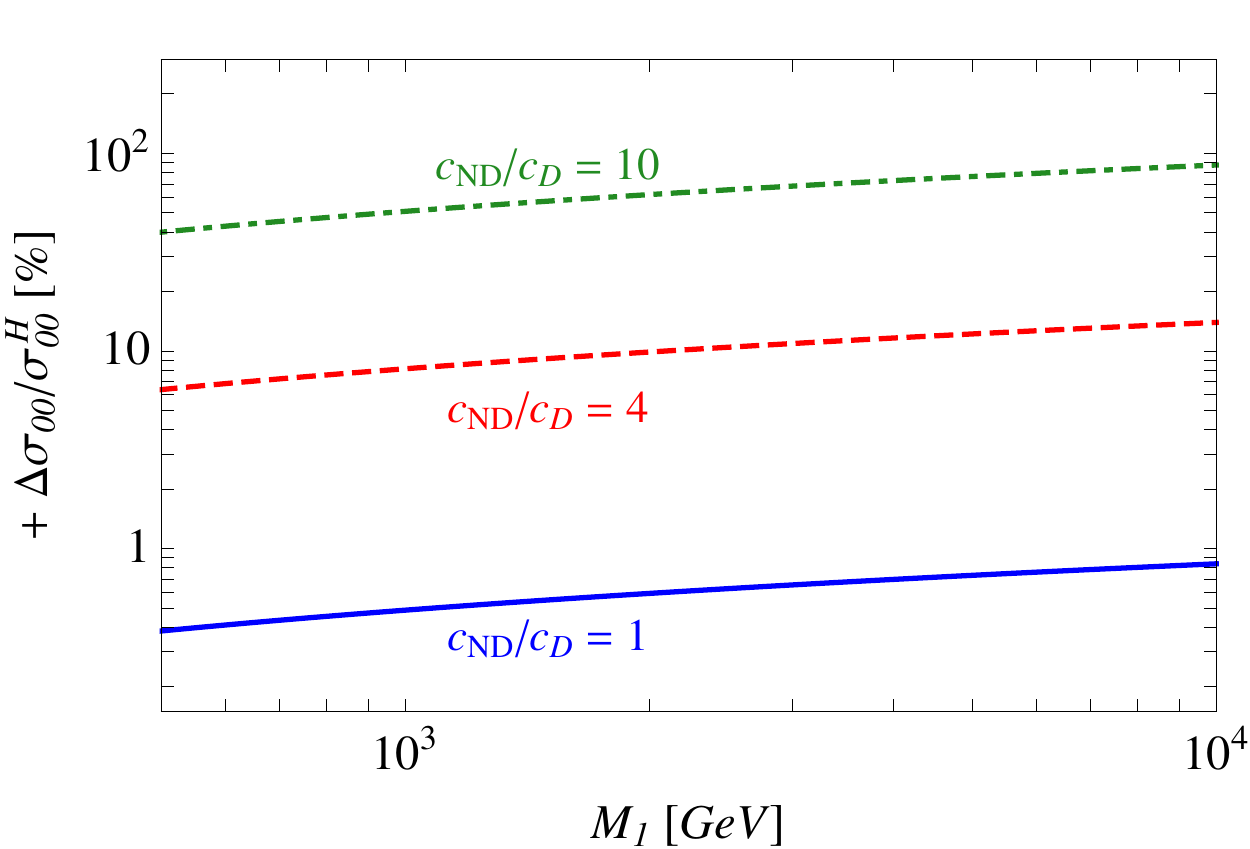}
\caption{\textit{BN-violating IR corrections to the annihilation cross section $\sigma_{00}$ defined in Eq.~(\ref{eq:DMannihilation}). We plot the relative correction $\Delta\sigma_{00}/\sigma_{00}^{\rm H}$ in Eq.~(\ref{eq:RelCorr}) as a function of the DM mass for three different values of the ratio $C_{\rm ND}/C_{\rm D}$.}}
\label{fig:BNviolation}
\end{center}
\end{figure}

Let us now start our discussion about the cancellation theorems elaborating  in more details the annihilation processes described by Eq.~(\ref{bncompact}).
In particular we first analyze the situation in which the average over the initial flavors is not performed,
 leading to the violation of the BN theorem \cite{violation, generic, ew}.
The annihilation of DM particles in the early universe analyzed in this paper represents a completely different physical situation; it is interesting, as a consequence, to better investigate how the BN violation manifests itself in this context.
As explained in Section~\ref{sec:KLNviolation}, the average over the initial flavors is not required whenever the thermal history of the decoupling prevents the inclusion of the co-annihilation processes. 
In order to match with the formalism elaborated in \cite{violation, generic}, we first define the following inclusive cross sections
\begin{eqnarray}
\sigma_{0+}&\equiv&\sigma(\chi_0\chi^+\to f_1\bar{f}_2)~,\\
\sigma_{0-}&\equiv&\sigma(\chi_0\chi^-\to f_2\bar{f}_1)~,\\
\sigma_{+-}&\equiv&\sigma(\chi^+\chi^-\to f_1\bar{f}_1)+\sigma(\chi^+\chi^-\to f_2\bar{f}_2)~,\\
\sigma_{00}&\equiv&\sigma(\chi_0\chi_0\to f_1\bar{f}_1)+\sigma(\chi_0\chi_0\to f_2\bar{f}_2)~,\label{eq:DMannihilation}
\end{eqnarray}
with $\chi_0\equiv \chi^3$, $\chi^{\pm}\equiv (\chi^1\mp i\chi^2)/\sqrt{2}$. Using Eq.~(\ref{lagrangian}) we find
\begin{eqnarray}
\sigma_{0\pm}^{\rm H}&=&\frac{C_{\rm ND}^2N_{C}^f s[48 M_{1}^2 +s(12+v^2)]}{6\pi \Lambda^4\sqrt{s^2(4+v^2)^2-256 M_{1}^4}}~,\\
\sigma_{+-}^{\rm H}&=& \frac{
N_C^f s\{
C_{\rm D}^2[s(12+v^2)-48M_{1}^2]+C_{\rm ND}^2[48M_{1}^2+s(12+v^2)]
\}
}{6\pi\Lambda^4 \sqrt{s^2(4+v^2)^2-256M_{1}^4}}~,\\
\sigma_{00}^{\rm H}&=& \frac{C_{\rm D}^2 N_{C}^f s[s(12+v^2)-48M_{1}^2]}{6\pi \Lambda^4 \sqrt{s^2(4+v^2)^2-256 M_{1}^4}}~,
\end{eqnarray}
where the superscript $^{\rm H}$ indicates that these inclusive hard  cross sections are evaluated at the tree level while for the square of  the total energy in the center of mass frame we have
$s=4M_{1}^2/(1-v^2/4)$. The color factor $N_{C}^f$ is equal to $3$ ($1$) for final state involving quarks (leptons).  
Considering the expansion $v\sigma_{ij}^{\rm H}=a_{ij}+b_{ij}v^2
+\mathcal{O}(v^4)$, and dropping terms $\mathcal{O}(v^4)$, we find
\begin{eqnarray}
v\sigma_{0\pm}^{\rm H}&=&\frac{2M_{1}^2C_{\rm ND}^2N_C^f}{\pi \Lambda^4}\left(2+\frac{v^2}{3}\right)~,\label{eq:sigmav0+}\\
v\sigma_{+-}^{\rm H}&=&\frac{2M_{1}^2N_C^f}{\pi\Lambda^4}\left[
2C_{\rm ND}^2+\frac{v^2(C_{\rm D}+C_{\rm ND}^2)}{3}
\right]~,\\
v\sigma_{00}^{\rm H}&=&\frac{2M_{1}^2C_{\rm D}^2N_C v^2}{3\pi\Lambda^4}~,\label{eq:sigmav00}
\end{eqnarray}
where in particular the p-wave behavior of $v\sigma_{00}^{\rm H}$ becomes evident.
Notice that these cross sections satisfy the sum rules dictated by the SU(2)$_L$ invariance  \cite{generic}, namely 
$\sigma_{++}=\sigma_{--}$, $\sigma_{0+}=\sigma_{0-}$, $\sigma_{00}=\sigma_{++}+\sigma_{+-}-\sigma_{0+}$, where in this particular case $\sigma_{\pm\pm}=0$.  The inclusion of the one-loop electroweak IR corrections changes the tree level cross section  $\sigma_{00}^{\rm H}$ into its ``dressed" form \cite{generic}
\begin{equation}\label{eq:BNviolation}
\sigma_{00}=\sigma_{00}^{\rm H}-4\;(\sigma_{+-}^{\rm H}-2\;\sigma_{3+}^{\rm H})\mathcal{L}_W~,
\end{equation}
with $\mathcal{L}_W$ given in Eq.~(\ref{eq:EikonalNonRel}). 
This extremely concise formula contains the sum of real and virtual IR corrections to the DM annihilation cross section,
and the logarithmic dependence is the trademark of the BN violation.
Defining $\Delta\sigma_{00}\equiv \sigma_{00}
-\sigma_{00}^{\rm H}$, and using for the DM relative velocity at the freeze-out the value $v^2\sim 1/4$ we find the relative correction
\begin{equation}\label{eq:RelCorr}
\frac{\Delta\sigma_{00}}{\sigma_{00}^{\rm H}}=
\left(
\frac{\sigma_{0+}^{\rm H}}{\sigma_{00}^{\rm H}}-1
\right)\frac{\alpha_W}{4\pi}\,\frac{4}{3}\,\ln\frac{M_1^2}{m_W^2}~.
\end{equation}
The BN-violating IR corrections always \textit{enhance} the tree level cross section $\sigma_{00}^{\rm H}$ if $\sigma_{0+}^{\rm H}\geqslant\sigma_{00}^{\rm H}$; considering the analytic expressions given in Eqs.~(\ref{eq:sigmav0+},\,\ref{eq:sigmav00}), this remains true as long as $C_{\rm ND}/C_{\rm D}\gtrsim 0.2$. 
We show the ratio $\Delta\sigma_{00}/\sigma_{00}^{\rm H}$  in Fig.~\ref{fig:BNviolation} as a function of the DM mass considering three different benchmark values for the coefficients of the effective Lagrangian in Eq.~(\ref{lagrangian}), $C_{\rm ND}/C_{\rm D}=1,4,10$. 
For a TeV-scale DM particle the effect of the 
BN-violating IR corrections can easily reach the $50\%$ level. 
Even in this simple example, therefore, we can fully appreciate the considerable impact of these corrections. Bearing in mind that the one-loop virtual corrections always lead to a suppression of the cross section, in fact, the inclusion of the real emission processes drastically overturns this scenario: the net effect of the radiative corrections becomes positive, increasing the  DM annihilation rate.  This shows that
the violation of the BN theorem may lead to corrections much larger than those analyzed, for instance, in Refs. 
\cite{Baro:2009na, Herrmann:2009mp,Boudjema:2011ig, Akcay:2012db,Harz:2012fz} for the supersymmetric DM and where  corrections to  the DM density of at most 10\% were found.

Given the astonishing precision reached for the experimental measurement of the physical density of cold DM in the universe ($\Omega_{\rm DM}h^2 = 0.1199\pm 0.0027$ \cite{Ade:2013zuv}), a careful treatment of these corrections is mandatory. 
Such corrections have to be added to those recently pointed out in Ref.~\cite{Steigman:2012nb}, originating from a proper 
consideration of the temperature dependence of the number of degrees of freedom,
which also can be of the order of 60\%.

We stress once again that, beyond the toy model used for illustrative  purposes in this paper, the only ingredient needed for the computation of these corrections is the isospin quantum number of the DM multiplet together with the eikonal integral in Eq.~(\ref{eq:EikonalNonRel}).

Let us now move to discuss in more details the situation in which the co-annihilation processes become important.
In this case the necessity to perform the average over the initial flavors washes out the logarithm in Eq.~(\ref{eq:BNviolation}) as a consequence of the BN theorem. Nevertheless, as explained in the previous Section, a residual logarithmic correction proportional to the mass splitting $\Delta_a$ survives. This correction, whose magnitude is estimated in Eq.~(\ref{eq:ResidualKLNviolation}), originates from the different Boltzmann weights that affect each component of the DM multiplet in the thermal bath, and formally violates  the BN theorem.
 
Moreover, the isospin-breaking terms that lift the degeneracy in the DM multiplet are by themselves a possible source of BN violation. Besides the corrections of dynamical type in Eq.~(\ref{eq:ResidualKLNviolation}), therefore, one should include also these isospin-breaking effects. However, it is possible to show that the latter are sub-leading if compared to the former. To be convinced, one can account for the mass splitting effects introducing the following improved
 eikonal current $J_{\rm eik,\,\Delta}^{\mu}$  (see \cite{anomalous})
\begin{equation}
J_{\rm eik}^{\mu} = g\,\frac{k^{\mu}}{p\cdot k}~~~~\Longrightarrow~~~~
J_{\rm eik,\,\Delta}^{\mu}\equiv
g\,\frac{(2k^{\mu} + M_1\Delta_a \gamma^{\mu})}{2p\cdot k}~,
\end{equation}
describing the emission of a soft  gauge boson with four-momentum $k$ from an initial DM particle with four-momentum $p$.
An explicit calculation shows that the residual logarithmic corrections arise from the diagrams involving only emission from the initial states,\footnote{In particular we find that the corrections proportional to the mass splitting coming from the interference between initial and final legs cancel out.} and that their explicit form has the following 
structure
\be\label{eq:ResidualKLNv2}
{\cal O}\left(\Delta_a^2\cdot{\cal N}\cdot\frac{\alpha_W}{4\pi}\cdot\ln\frac{M_1^2}{m_W^2}
\right)~,
\ee
thus producing a sub-dominant effect if compared to the thermal correction in Eq.~(\ref{eq:ResidualKLNviolation}).

Let us now close our discussion with a final comment. 
Considering the annihilation of DM particles, the Sommerfeld enhancement might play a relevant role.  In a nutshell the Sommerfeld enhancement
is a non-relativistic effect generated by the exchange of light force carriers between the incoming DM particles, resulting in the possibility to form a resonant bound state. Using a non-relativistic resummation technique, it has been shown \cite{Hisano:2002fk}
 that this enhancement can affect also the calculation of the relic abundance, leading
  in some cases to a reduction of 50\% (see also \cite{Hryczuk:2010zi, iengo}). However, relying on the creation of a bound state, 
  the Sommerfeld enhancement must fulfill two special conditions in order to reach such large values. Describing
   the exchange of  massive W bosons with a Yukawa potential,\footnote{In order to allow a clear and immediate comparison with the situation analyzed in this paper, we focus our discussion
    on the case in which the Sommerfeld enhancement is primed by the exchange of SM gauge bosons rather than new dark mediators, see Ref. \cite{ArkaniHamed:2008qn}.}
  the first condition is that the range of this interaction is larger than the typical Bohr radius of the system 
  consisting of the two DM incoming particles. This request is achieved whenever $\alpha_WM_1 \gtrsim m_W$,
  namely, being $\alpha_W\approx 0.03$, for a TeV-scale DM particle.
  More importantly,  the characteristic energy of the interaction must be larger that the kinetic DM energy. This second prerogative leads to the condition $v \lesssim \alpha_W$. This inequality 
  can be easily satisfied in the present-day annihilations  ($v\sim 10^{-3}$), but it seems very unlikely in the early universe ($v\sim 1/2$). The large corrections quoted above, in fact, 
  can happen only in presence of a state slightly heavier than the DM particle. 
  This state can therefore be produced by the incoming DM particles with almost zero velocity, allowing the possibility to generate the Sommerfeld enhancement. More quantitatively, this condition reads $2\Delta_a \lesssim \alpha_W^2 + v^2/4$, where the r.h.s. follows from the sum of the characteristic Bohr energy of the potential and the collision energy of the two DM particles. Using $v\sim 1/2$ we have $\Delta_a \lesssim 1/30$. As noticed in Section~\ref{sec:EWfreeze}, the reader should also keep in mind that, being a virtual effect, the Sommerfeld enhancement is always proportional to $|C_{\rm D}|^2$.
  Notice also that in the case of scalar DM, the operators controlling the mass splitting
within the multiplet are generically of lower dimension than for fermion DM, so one expects
 $\Delta_a$ to be larger for  scalar DM than for fermion DM.
 All in all, a complementarity between the Sommerfeld enhancement and the IR corrections arises. On the one hand, in presence of a small mass splitting between the component of the DM multiplet, the IR corrections are screened by the BN theorem thus affecting the Sommerfeld enhancement only with the residual BN-violating log in Eq. (\ref{eq:ResidualKLNviolation}). On the other hand, we argue that in presence of a large mass splitting 
  the BN-violating IR corrections described in this paper will be the dominant effect.

\section{Conclusions}
\label{sec:conclusions}

In this paper we have discussed the important role of EW corrections 
for the calculation of the DM relic abundance. 
In fact, the cancellation theorems which prevent log corrections
to appear when computing annihilation cross sections with the inclusion
of gauge boson emission, may be evaded.
 
In the situation described as follows
\begin{itemize}
\item the DM particle is the electrically neutral component of a SU(2)$_L$ mutliplet,
whose components are not degenerate in mass;
\item the DM mass $M_1$ is much heavier than the weak scale $M_1\gg m_W$;
\item the relative mass splitting within the multiplet 
is sufficiently large $\Delta_a=(M_a-M_1)/M_1\gtrsim 1/25$
\end{itemize}
we have found that the cancellation theorems  are not operative and there are potentially
large corrections to the DM relic abundance.

For instance, for a TeV-scale DM particle the effect of the 
BN-violating IR corrections can easily reach an \textit{enhancement} of the $50\%$ level (see Fig.~\ref{fig:BNviolation}).
These corrections have to be added to those due to the entropy changes not considered in this work, {\it e.g.} due to the QCD transition from deconfinement to confinement, mass-generation
above the electroweak scale, and other possible transitions recently analyzed in Ref. [23], in which the change in the number of degrees of freedom give corrections that can be of the same order of the ones considered here.
Therefore, in the era of precision cosmology where the energy density of
cold DM in the universe is measured at the percent level,
reliable calculations of the DM relic abundance need to take these effects into account.


\section*{Acknowledgments}

ADS acknowledges partial support from the  European Union FP7  ITN INVISIBLES (Marie Curie Actions, PITN-GA-2011-289442).
The work of AU is supported by the ERC Advanced Grant n$^{\circ}$ $267985$, ``Electroweak Symmetry Breaking, Flavour and Dark Matter: One Solution for Three Mysteries" (DaMeSyFla). The work of AU is dedicated to Francesco Caracciolo.


\appendix

\section{Cancellation of divergences for mass-degenerate states}
\label{app:CollinearCancellation}

In this Appendix we explicitly work out the calculation of the cancellation of the collinear divergent logarithms, to show how the cancellation theorems work in the case where
the DM belongs to a multiplet whose states are degenerate in mass.
For simplicity, we consider only the non diagonal term in the Lagrangian of Eq.~(\ref{lagrangian}):
\be
 \mathscr{L}_{\rm eff,\,ND}=i\frac{C_{\rm  ND}}{\Lambda^2}\epsilon_{abc}\left(\overline{L}\gamma_\mu P_L \sigma^cL\right)
\left(\overline{\chi}^a\gamma^\mu  \chi^b\right)~.
\ee
Collinear divergences arise when the scalar product $k\cdot p$ goes to $0$, where $k$ is the momentum of the $W$ emitted or exchanged in the loop, and $p$ is the momentum of one of the two final state leptons.
In the non-relativistic limit for the initial state DM particles we expect no collinear divergence. Indeed, if $k_1$ is the momentum of the DM particle and $M$ is the common mass of the DM multiplet, we have
\be
k_1^\mu\simeq(M,0,0,k_1)~, \qquad k^\mu \simeq (k,k\sin\theta,0,k\cos\theta)~,
\ee
with $M\gg k_1,k$. The scalar product is then
\be
k_1\cdot k \simeq Mk+k_1k\cos\theta~,
\ee
that if $k\neq0$ is non zero for every value of the angle $\theta$.

\begin{figure}[t!]
\begin{center}
\includegraphics[scale=1]{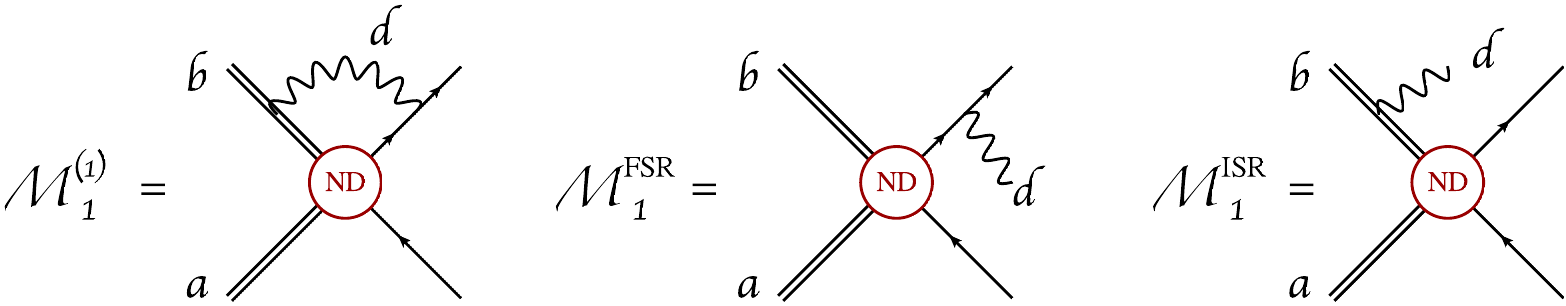}
\caption{{\small\textit{Feynman diagrams corresponding to the scattering amplitudes in Eqs.~(\ref{eq:Virtual}-\ref{eq:FSR}).}}}
\label{fig:Diagrams}
\end{center}
\end{figure}

For what concerns the emission of a $W$ in the final state, we already know that the divergences coming from this effect are cancelled by  those coming from the exchange of a $W$ between the two final state legs.
Since, as we showed, there is no collinear divergence from initial state radiation, the only non trivial point is the cancellation of divergences in the interference between initial and final state radiation.
To do this, we define the three amplitudes
\begin{eqnarray}
\MLNDi &\equiv& \frac{g^2C_{\rm ND}}{2\Lambda^2}\epsilon_{aec}\epsilon_{bdc}\int\frac{{\rm d}^4k}{(2\pi)^4}\left\{
\frac{(g_{\alpha\beta}-k_{\alpha}k_{\beta}/m^2)}{(k^2-m^2)}
\bar{v}(k_2)\gamma^{\mu}\frac{(\slashed{k}_1-\slashed{k}+M)}{[(k_1-k)^2-M^2]}\gamma^{\alpha}u(k_1)
\right.\nonumber  \\
&&\left.
\times~\bar{u}(p_1)\gamma^{\beta}\frac{(\slashed{p}_1-\slashed{k})}{(-p_1\cdot k+m^2)}
\gamma_{\mu}\sigma^d\sigma^eP_Lv(p_2)
\right\}~,\label{eq:Virtual}\\
\MINDi  &\equiv& -\frac{igC_{\rm ND}}{\Lambda^2}\epsilon_{ace}\epsilon_{cdb}\bar{v}(k_2)
\gamma^{\mu}\frac{(\slashed{k}_1-\slashed{k}+M)}{[(k_1-k)^2-M^2]}
\slashed{\epsilon}^*(k)u(k_1)\bar{u}(p_1)\gamma_{\mu}P_L\sigma^ev(p_2)~,  \\
\MFNDi &\equiv&  \frac{gC_{\rm ND}}{2\Lambda^2}
\bar{v}(k_2)\gamma^{\mu}u(k_1)\bar{u}(p_1)\slashed{\epsilon}^*(k)
\frac{(\slashed{p}_1+\slashed{k})}{(2p_1\cdot k+m^2)}
\gamma_{\mu}[\epsilon_{abd}+i\delta_{bd}\sigma^a-i\delta_{ad}\sigma^b]P_Lv(p_2)~,\nonumber\\
&&\label{eq:FSR}
\end{eqnarray}
where $m$ is the $W$ mass. The corresponding Feynman diagrams are shown in Fig.~\ref{fig:Diagrams}.
Because of the Lorentz structure of the diagrams, to show that the cancellation holds,  it is sufficient to show that
\be
\MTND\cdot{\left(\MLNDi\right)}^* + \MFNDi\cdot{\left(\MINDi\right)}^*
\ee
has no logarithmic collinear divergences. The divergences coming from the other similar diagrams will cancel in the same way.
In the rest of this appendix we compute separately the two terms using the collinear approximation and show that their sum is indeed finite.

\subsection{The amplitude of the $2\rightarrow 3$ process}
The amplitude of the $2\rightarrow 3$ process reads
\begin{eqnarray}
\MFNDi\cdot{\left(\MINDi\right)}^* & = & -2ig^2\frac{C_{\rm ND}^2}{\Lambda^4} \epsilon^*_\alpha\epsilon_\beta \spinorvbar{k_2}\gamma^\mu\spinoru{k_1} \nonumber \\
&\times&
\spinorubar{p_1}\gamma^\alpha\frac{\slashed{p_1}+\slashed{k}}{2p_1\cdot k+m^2}\gamma_\mu\left[\epsilon_{abd}+i\epsilon_{abf}\epsilon_{gdf}\sigma^g\right]P_L\spinorv{p_2} \nonumber \\
&\times&\epsilon_{aec}\epsilon_{bdc}\spinorvbar{p_2}\sigma^e\gamma_\nu P_L\spinoru{p_1}\spinorubar{k_1}\gamma^\beta\frac{\slashed{k}_1-\slashed{k}+M}{(k_1-k)^2-M^2}\gamma^\nu\spinorv{k_2}~.
\label{eqFSRNDiISRNDi}
\end{eqnarray}
The $\text{SU}(2)_L$ indices give
\begin{eqnarray}
\sum_{c}\epsilon_{abd}\epsilon_{aec}\epsilon_{bdc} & = & \epsilon_{abd} (\delta_{ab}\delta_{de}-\delta_{ad}\delta_{be}) = 0~, \\
\sum_{c,e,d,f} \epsilon_{abf}\epsilon_{gdf}\epsilon_{aec}\epsilon_{bdc}\sigma^g\sigma^e & = &
(\delta_{ag}\delta_{bd}-\delta_{ad}\delta_{bg})(\delta_{ab}\delta_{de}-\delta_{ad}\delta_{be})\sigma^g\sigma^e \nonumber \\
& = & \delta_{ab}\delta_{bd}\sigma^a\sigma^d - \delta_{ab}\delta_{ad}\sigma^b\sigma^d - \delta_{ad}\delta_{bd}\sigma^a\sigma^d + \delta_{ad}\sigma^b\sigma^b \nonumber \\
& = & - \delta_{ad}\delta_{bd}\sigma^a\sigma^d + \delta_{ad}\sigma^b\sigma^b = 
\left\{
\begin{array}{l} 0 \text{ when } a=b=d \\ 1 \text{ when } a=d\neq b~, \end{array}
\right.
\end{eqnarray}
so that (\ref{eqFSRNDiISRNDi}) does not vanish only for $a=d\neq b$, and in this case we obtain
\begin{eqnarray}
\MFNDi\cdot{\left(\MINDi\right)}^* & = &
2g^2\frac{C_{\rm ND}^2}{\Lambda^4} \epsilon^*_\alpha\epsilon_\beta \spinorvbar{k_2}\gamma^\mu\spinoru{k_1}\spinorubar{p_1}\gamma^\alpha\frac{\slashed{p_1}+\slashed{k}}{2p_1\cdot k+m^2}\gamma_\mu P_L\spinorv{p_2} \nonumber \\
&\times&
\spinorvbar{p_2}\gamma_\nu P_L\spinoru{p_1}\spinorubar{k_1}\gamma^\beta\frac{\slashed{k}_1-\slashed{k}+M}{(k_1-k)^2-M^2}\gamma^\nu\spinorv{k_2}.
\end{eqnarray}
Summing over spin and polarization states we get
\begin{eqnarray}
\Big<\MFNDi\cdot{\left(\MINDi\right)}^*\Big> &=& -\frac{g^2}{2}\frac{C_{\rm ND}^2}{\Lambda^4} \frac{1}{2p_1\cdot k+m^2}\frac{1}{(k_1-k)^2-M^2} \left(g_{\alpha\beta}-\frac{k_\alpha k_\beta}{m^2}\right) \nonumber \\
&\times& \tr\left[\gamma^\mu(\slashed{k}_1+M)\gamma^\beta(\slashed{k}_1-\slashed{k}+M)\gamma^\nu(\slashed{k}_2-M)\right]  \nonumber \\
&\times&
\tr\left[\slashed{p}_1\gamma^\alpha(\slashed{p}_1+\slashed{k})\gamma_\mu\slashed{p}_2\gamma_\nu P_L\right]~.
\label{eqWemissioninterference}
\end{eqnarray}
To compute the cross section we have to integrate Eq. (\ref{eqWemissioninterference}) with the measure
\begin{equation}
\frac{\d^3p_1}{(2\pi)^32E_1}\frac{\d^3p_2}{(2\pi)^32E_2}\frac{\d^3k}{(2\pi)^32E_k}\delta^4(P_\text{in}-p_1-p_2-k)~,
\label{eqmeasure}
\end{equation}
where $P_\text{in}$ is the total momentum of the initial state.
We want to factorize the $\d^3k$ integral using the collinear approximation. In order to do so, we multiply the measure in Eq.~(\ref{eqmeasure}) by a factor of $1=\int{\d^4p\,\delta^4(p-p_1-k)}$, so that we have
\begin{equation}
\int \frac{\d^3p_2}{(2\pi)^32E_2}\frac{\d^3p_1}{(2\pi)^32E_1}\frac{\d^3k}{(2\pi)^32E_k}
\left[
\int \d p^0\delta(p^0-p_1^0-k^0) \d^3p \delta^3(\vec{p}-\vec{p}_1-\vec{k})
\right]
\delta^4(P_\text{in}-p_1-p_2-k)~.
\end{equation}
Now we integrate over $\d p^0$ and over $\d^3p_1$ using the first two Dirac deltas to get
\begin{equation}
\int \frac{\d^3p_2}{(2\pi)^32E_2}\frac{\d^3k}{(2\pi)^32E_k} \frac{\d^3p}{(2\pi)^32|\vec{p}-\vec{k}|}
\delta^4(P_\text{in}-p_1-p_2-k)~.
\end{equation}
Finally using the collinear approximation $\vec{k}\approx x\vec{p}$ we obtain the factorized measure
\begin{equation}
\int \frac{\d^3p}{(2\pi)^32E}\frac{\d^3p_2}{(2\pi)^32E_2}\delta^4(P_\text{in}-p-p_2)\frac{\d^3k}{(2\pi)^32E_k} \left(\frac{1}{1-x}\right)~.
\end{equation}
The $\d^3k$ integral can be written in spherical coordinates, using the collinear approximation, as
\begin{equation}
\int{\frac{\d^3k}{(2\pi)^32E_k}} = \int{\frac{|\vec{k}|^2{\rm d}k \sin\theta \d\theta \d\phi}{(2\pi)^32E_k}}
\approx
\frac{1}{2}\int_0^1{\d x x}\int_{-1}^{+1}{\d(\cos\theta)}\int_0^{2\pi}{\d\phi}~.
\end{equation}
In the collinear limit the traces become
\begin{equation}
(1-x)\left(g_{\alpha\beta}-x^2\frac{p_\alpha p_\beta}{m^2}\right)
\tr\left[\gamma^\mu(\slashed{k}_1+M)\gamma^\beta(\slashed{k}_1-x\slashed{p}+M)\gamma^\nu(\slashed{k}_2-M)\right]
\tr\left[\slashed{p}\gamma^\alpha(\slashed{p})\gamma_\mu\slashed{p}_2\gamma_\nu P_L\right], 
\end{equation}
which, using $p^2=0$, reduces to
\begin{equation}
64(1-x)k_1\cdot p\left[(k_1\cdot p)(k_2\cdot p_2) + (k_1\cdot p_2)(k_2\cdot p) + M^2p\cdot p_2\right]~.
\label{eqtracesresult}
\end{equation}
In the non-relativistic limit for the two DM particles, we can take $k_1=k_2=(M,0,0,0)$. Then $p=(M,0,0,M)$ and $p_2=(M,0,0,-M)$, and  Eq. (\ref{eqtracesresult}) gives simply $256(1-x)M^6$.
We have then
\begin{eqnarray}
\Big<\MFNDi\cdot{\left(\MINDi\right)}^*\Big> & \leadsto & -16g^2\frac{C_{\rm ND}^2}{\Lambda^4}\frac{M^8}{(2\pi)^3}  \\
&\times&
\int_0^{2\pi}{\d\phi}\int_{-1}^{+1}{\d(\cos\theta)}
\int_0^1{\d x \frac{x}{(2p_1\cdot k + m^2)[(k_1-k)^2-M^2]} }~, \nonumber
\end{eqnarray}
where with the symbol $\leadsto$ we mean that we consider only the collinear divergent part.

The three vectors $k$, $p_1$ and $p_2$ can be parametrized as
\begin{eqnarray}
k & = & \left(xE,\sqrt{x^2E^2-m^2}\sin\theta\cos\phi,\sqrt{x^2E^2-m^2}
\sin\theta\sin\phi,\sqrt{x^2E^2-m^2}\cos\theta\right)~, \\
p_1 & = & \left(yE,-\sqrt{x^2E^2-m^2}\sin\theta\cos\phi,-\sqrt{x^2E^2-m
^2}\sin\theta\sin\phi,\sqrt{y^2E^2-(x^2E^2-m^2)\sin^2\theta}\right)~, \nonumber \\ \\
p_2 & = & \left( zE,0,0,-zE \right)~.
\end{eqnarray}
Ignoring the $W$ mass, at order $\theta^2$ the scalar product $k\cdot p_1$ is equal to
\begin{equation}
k\cdot p_1 \simeq \frac{x}{2(1-x)}E^2\theta^2~.
\end{equation}
We substitute in the integral the expressions
\begin{eqnarray}
2p_1\cdot k + m^2 & \longrightarrow & \frac{x}{1-x}M^2\theta^2 + m^2~, \\
(k_1-k)^2-M^2 & \longrightarrow & 2xM^2 + m^2~, \\
\int \d(\cos\theta) & \longrightarrow & \int \theta \d\theta~,
\end{eqnarray}
so that
\begin{equation}
\Big<\MFNDi\cdot{\left(\MINDi\right)}^*\Big>\leadsto -64g^2\frac{C_{\rm ND}^2}{\Lambda^4}\frac{M^8}{(2\pi)^2}\int_0^1{\d x}
\int_{0}^{2\pi}{\d\theta \frac{\theta x}{\left(\frac{x}{1-x}M^2\theta^2 + m
^2\right)\left[2xM^2 + m^2\right]} }~.
\end{equation}
The $\d\theta$ integral is
\begin{equation}
\frac{1}{2}\int_0^{2\pi}{\frac{\d\theta^2}{\frac{x}{1-x}M^2\theta^2 + m^2}} =
\frac{1-x}{2xM^2}\log\left(1+\pi^2\frac{x}{1-x}\frac{M^2}{m^2}\right)
\end{equation}
and therefore, keeping only the collinear log-divergent part, we obtain
\begin{equation}
\langle\MFNDi\cdot{\left(\MINDi\right)}^*\rangle\leadsto -32g^2\frac{C_{\rm ND}^2}{\Lambda^4}\frac{M^6}{(2\pi)^2}
\left(
\int_{m/M}^1{\d x
\frac{1-x}{2xM^2+m^2} }\right)
\log\left(\frac{M^2}{m^2}\right)~.
\label{eqlogloop}
\end{equation}

\subsection{The amplitude of the $2\rightarrow 2$ process}
The amplitude of the $2\rightarrow 2$ process reads

\begin{eqnarray}
\MTND\cdot{\left(\MLNDi\right)}^* & = & -2g^2\frac{C_{\rm ND}^2}{\Lambda^4} \epsilon_{abf}\epsilon_{aec}\epsilon_{bdc}
\spinorvbar{k_2} \gamma^\nu \spinoru{k_1}\spinorubar{p_1}\gamma_\nu P_L\sigma^f\spinorv{p_2} \nonumber \\
&\times& \left(\int\frac{{\rm d}^4k}{(2\pi)^4} \frac{g_{\alpha\beta}-\frac{k_\alpha k_\beta}{m^2}}{k^2-m^2+i\epsilon}
\spinorvbar{k_2}\gamma^\mu\frac{\slashed{k}_1-\slashed{k}+M}{(k_1-k)^2-M^2+i\epsilon}\gamma^\alpha\spinoru{k_1} \right. \nonumber \\
&\times& {\left. \spinorubar{p_1}\gamma^\beta\frac{\slashed{p}_1-\slashed{k}}{(p_1-k)^2}\gamma_\mu\sigma^d\sigma^eP_L\spinorv{p_2} \right)}^*~.
\end{eqnarray}
In $\text{SU}(2)_L$ space we have
\begin{equation}
\epsilon_{aec}\epsilon_{bdc}\epsilon_{abf} = (\delta_{ab}\delta_{de}-\delta_{ad}\delta_{bc})\epsilon_{abf} = -\delta_{ad}\delta_{bc}\epsilon_{abf}~,
\end{equation}
that is non-zero  only for $a=d\neq b$; therefore we get
\begin{eqnarray}
\MTND\cdot{\left(\MLNDi\right)}^* & = & +2g^2\frac{C_{\rm ND}^2}{\Lambda^4} \epsilon_{abf}  \nonumber \\
&\times& \int\frac{{\rm d}^4k}{(2\pi)^4} \left(g_{\alpha\beta}-\frac{k_\alpha k_\beta}{m^2}\right)
\frac{1}{k^2-m^2-i\epsilon}\frac{1}{(k_1-k)^2-M^2-i\epsilon}\frac{1}{(p_1-k)^2-i\epsilon} 
\nonumber \\
&\times& \spinorvbar{k_2} \gamma^\nu \spinoru{k_1}
\spinorubar{k_1}\gamma^\alpha(\slashed{k}_1-\slashed{k}+M)\gamma^\mu\spinorv{k_2}
\spinorubar{p_1}\gamma_\nu P_L\sigma^f\spinorv{p_2}  \nonumber \\
& \times & \spinorvbar{p_2}\gamma_\mu(\slashed{p}_1-\slashed{k})\gamma^\beta P_L\sigma^b\sigma^a\spinoru{p_1}~.
\label{eqloopsigma}
\end{eqnarray}
Since $a$, $b$ are different, $\sigma^b\sigma^a = -i\epsilon_{abc}\sigma^c$. There is only one possible value for $c$ and  $f$  must be equal to it. We can therefore write $\sigma^b\sigma^a$ with $-i\epsilon_{abf}\sigma^f$ (without sum over $f$); this multiplied by the other $\epsilon_{abf}$ gives $(\epsilon_{abf})^2=1$. In the $\text{SU}(2)_L$ space we have
\begin{equation}
\spinorvbar{p_2}_i\sigma^f_{ij}\spinoru{p_1}_j\spinorubar{p_1}_j\sigma^f_{ji}\spinorv{p_2}_i~, \qquad\text{no summation over indices},
\end{equation}
where the value of the indices $i,j$ is fixed by the final state. Since $\sigma^f_{ij}\sigma^f_{ji} = 1$ we can simply forget about the $\sigma$ matrices in Eq.~(\ref{eqloopsigma}).
We have then (after summing over spin states)
\begin{eqnarray}
\Big<\MTND\cdot{\left(\MLNDi\right)}^*\Big> & = & -\frac{i}{2}g^2\frac{C_{\rm ND}^2}{\Lambda^4}
\int\frac{{\rm d}^4k}{(2\pi)^4}\left(g_{\alpha\beta}-\frac{k_\alpha k_\beta}{m^2}\right)  \nonumber \\
&\times&
\frac{1}{k^2-m^2-i\epsilon}\frac{1}{(k_1-k)^2-M^2-i\epsilon}\frac{1}{(p_1-k)^2-i\epsilon} \nonumber \\
& \times &
\tr\left[\gamma^\mu(\slashed{k}_1+M)\gamma^\beta(\slashed{k}_1-\slashed{k}+M)\gamma^\nu(\slashed{k}_2-M)\right]  \nonumber \\
& \times &
\tr\left[\slashed{p}_1\gamma_\nu\slashed{p}_2\gamma_\mu(\slashed{p}_1-\slashed{k})\gamma^\beta P_L\right]~.
\end{eqnarray}
We now have to perform the $\d k^0$ integral using the Cauchy formula. The three factors in the denominator coming from the three propagators give six poles in the complex plane. Since we are interested only in the part that is divergent for $p_1\cdot k\rightarrow 0$, we can consider only the poles coming from the propagator of the $W$, ignoring the others. Closing the contour of integration in the upper or in the lower semi-plane, we get 
\begin{equation}
\int{\frac{\d^4k}{(2\pi)^4}\frac{1}{k^2-m^2-i\epsilon}} \stackrel{\epsilon\rightarrow0}{=} i\int{\frac{\d^3k}{(2\pi)^3}\frac{1}{2E_k}}~,
\end{equation}
so that
\begin{eqnarray}
\Big<\MTND\cdot{\left(\MLNDi\right)}^*\Big> & \approx & \frac{1}{2}g^2\frac{C_{\rm ND}^2}{\Lambda^4}
\int\frac{\d^3k}{(2\pi)^3}\left(g_{\alpha\beta}-\frac{k_\alpha k_\beta}{m^2}\right)
\frac{1}{2E_k}\frac{1}{(k_1-k)^2-M^2}\frac{1}{(p_1-k)^2} \times \nonumber \\
&\times& \tr\left[\gamma^\mu(\slashed{k}_1+M)\gamma^\beta(\slashed{k}_1-\slashed{k}+M)\gamma^\nu(\slashed{k}_2-M)\right] \nonumber \\
&\times&
\tr\left[\slashed{p}_1\gamma_\nu\slashed{p}_2\gamma_\mu(\slashed{p}_1-\slashed{k})\gamma^\beta P_L\right].
\end{eqnarray}
If we define $x$ as in the previous subsection, in the collinear limit we have $k\approx xp_1$ with $p_1\approx(E,0,0,E)$ and $p_2\approx(E,0,0,-E)$ where $2E$ is the center of mass energy.
The traces become then
\begin{equation}
(1-x)\left(g_{\alpha\beta}-x^2\frac{p_{1\alpha} p_{1\beta}}{m^2}\right)
\tr\left[\gamma^\mu(\slashed{k}_1+M)\gamma^\beta(\slashed{k}_1-x\slashed{p}_1+M)\gamma^\nu(\slashed{k}_2-M)\right]
\tr\left[\slashed{p}_1\gamma^\alpha(\slashed{p}_1)\gamma_\mu\slashed{p}_2\gamma_\nu P_L\right]~,
\end{equation}
that is the same expression obtained in the previous subsection. In the non-relativistic limit we get  simply $256(1-x)M^6$.

The integral in $\d^3k$ can be written in spherical coordinates in the same way as before. Notice that, despite the fact that in a loop integral one should integrate over alla momenta up to $+\infty$, here the $\d x$ integral goes from $m/M$ to $1$. Physically this is due to the fact that, integrating around the $W$ propagator poles, we are putting the $W$ on-shell. Then this $W$ behaves like a physical particle, and its momenta cannot be greater than  the initial energy.
The final result is then
\begin{equation}
\Big<\MTND\cdot{\left(\MLNDi\right)}^*\Big> \leadsto 64g^2\frac{C_{\rm ND}^2}{\Lambda^4}\frac{M^8}{(2\pi)^3}\int_0^{2\pi}{\d\phi}\int_{-1}^{+1}{\d(\cos\theta)}
\int_0^{1}{\d x \frac{x(1-x)}{(2p_1\cdot k + m^2)[(k_1-k)^2-M^2]} }~.
\end{equation}
In function of $x$, $\theta$ and $\phi$, $k$ is parametrized as
\begin{equation}
k = \left(xE,\sqrt{x^2E^2-m^2}\sin\theta\cos\phi,\sqrt{x^2E^2-m^2}\sin\theta\cos\phi,\sqrt{x^2E^2-m^2}\cos\theta\right)~.
\end{equation}
Ignoring the $W$ mass, at order $\theta^2$ the scalar product $k\cdot p_1$ becomes
\begin{equation}
p_1\cdot k = \frac{1}{2}xE^2\theta^2~,
\end{equation}
so that
\begin{equation}
\Big<\MTND\cdot{\left(\MLNDi\right)}^*\Big> \leadsto 64g^2\frac{C_{\rm ND}^2}{\Lambda^4}\frac{M^8}{(2\pi)^2}  \int_{m/M}^{1}{\d x}
\int_0^{2\pi}{ \d\theta \frac{\theta x(1-x)}{(xM^2\theta^2+m^2)(2xM^2+m^2)} }~.
\end{equation}
Performing the $\d\theta$ integral we obtain
\begin{equation}
\Big<\MTND\cdot{\left(\MLNDi\right)}^*\Big> \leadsto 32g^2\frac{C_{\rm ND}^2}{\Lambda^4}\frac{M^6}{(2\pi)^2}  \int_{m/M}^{1}{\d x \left\{ \frac{1-x}{2xM^2+m^2}\log\left(1+\pi^2x\frac{M^2}{m^2}\right) \right\} }~.
\end{equation}
The collinear log-divergent part is
\begin{equation}
\Big<\MFNDi\cdot{\left(\MINDi\right)}^*\Big>\leadsto 32g^2\frac{C_{\rm ND}^2}{\Lambda^4}\frac{M^6}{(2\pi)^2}
\left(
\int_{m/M}^{1}{\d x
\frac{1-x}{2xM^2+m^2} }\right)
\log\left(\frac{M^2}{m^2}\right)
\label{eqlogemission}
\end{equation}
that exactly cancels with (\ref{eqlogloop}).

\bibliographystyle{JHEP}


\end{document}